\documentstyle[times,epsfig,astrobib,useAMS]{mn2e}

\def\beb{\begin{thebibliography}{999}}
\def\eeb{\end{thebibliography}}
\def\bei{\begin{itemize}}
\def\eei{\end{itemize}}
\def\bef{\begin{figure}}
\def\eef{\end{figure}}
\def\ben{\begin{enumerate}}
\def\een{\end{enumerate}}
\def\beq{\begin{equation}}
\def\eeq{\end{equation}}
\def\ber{\begin{eqnarray}}
\def\eer{\end{eqnarray}}

\def\ab{{\bf A}}
\def\bb{{\bf B}}
\def\nb{{\bf \nabla}}
\def\rb{{\bf r}}

\newcommand{\msun}{\mbox{{\rm M}$_{\odot}$}}

\newcommand{\pdot}{\mbox{$\dot{P}$}}

\newcommand{\lsim}{\raisebox{-0.3ex}{\mbox{$\stackrel{<}{_\sim} \,$}}}

\title[{\it {Strange Pulsar}}]{Strange Pulsar Hypothesis}
\author[Ray Mandal et al.]
{Raka Dona Ray Mandal$^{1}$, 
Monika Sinha$^{1*}$, 
Manjari Bagchi $^{1}$, 
Sushan Konar$^{2}$, 
Mira Dey $^{1**}$
\and and Jishnu Dey $^{1}$\\ 
$^{1}$ Physics, Presidency College, Kolkata, India \\
$^{2}$ CTS \& Physics, Indian Institute of Technology, 
       Kharagpur, India \\
$*$ Research Fellow, CSIR, GOI \\ 
$**$ Emeritus Scientist, CSIR, GOI \\
e-mail : rakadona@vsnl.net,
monika2003@vsnl.net,
mnj2003@vsnl.net,
sushan@iitkgp.ac.in,
deyjm@giascl01.vsnl.net.in}

\begin{document}

\date{}

\pagerange{\pageref{firstpage}--\pageref{lastpage}} \pubyear{2005}

\maketitle

\label{firstpage}

\begin{abstract}
It  appears that there  is a  genuine shortage  of radio  pulsars with
surface magnetic fields  significantly smaller than $\sim 10^8$~Gauss.
We propose that the pulsars with very low magnetic fields are actually
strange stars locked  in a state of minimum  free energy and therefore
at a limiting value of the  magnetic field which can not be lowered by
the system spontaneously.
\end{abstract}


\section{Introduction}

A radio  pulsar is a  strongly magnetized rotating neutron  star.  The
dipolar  component of  the magnetic  field can  be estimated  from its
spin-period ($P$) and the period derivative ($\pdot$), apart from some
structural  constants (${\cal  B}  \propto \sqrt{P\pdot}$).   Magnetic
fields estimated in this  fashion broadly classifies the radio pulsars
in two categories - a)  isolated pulsars with rotation periods usually
above $1s$ and very  strong magnetic fields ($10^{11}-10^{13}$~G); (b)
binary/millisecond  pulsars  with much  shorter  rotation periods  and
considerably    weaker     magnetic    fields    ($10^{8}-10^{10}$~G).
Observations suggest  a connection between  the low magnetic  field of
the  second  group  with  their  being processed  in  binary  systems,
indicating  an  accretion-induced  field  decay  in  such  cases  (see
Bhattacharya~\citeyear{db02} for a review).

However, none  of the existing  mechanisms of field  evolution implies
that  the   final  field  should   saturate  to  a   particular  value
irrespective  of   the  evolutionary   history  of  a   given  pulsar.
Surprisingly,  the observed pulsar  population seems  to have  a lower
bound for the magnetic field strength.  Till date some fifteen-hundred
radio pulsars  have been discovered, for  a large number  of which the
values of  $P$ and $\pdot$  (and therefore ${\cal B}$)  are available.
Fig.[\ref{fig01}]  shows a  plot of  the magnetic  field vs.  the spin
period (${\cal B}-P$) of these  pulsars. It appears that there exist a
{\em minimum} value of the magnetic field, around $\sim 10^8$~G.

In the  present work, we try  to offer an explanation  for the minimum
magnetic field  by associating the  millisecond pulsars with  the {\em
strange stars}.  The relation between  the decay of the magnetic field
and the binary history of a  pulsar has been firmly established. It is
also  understood that the  very low-field  pulsars, in  particular the
millisecond  pulsars, are  products of  low-mass X-ray  binary systems
(LMXB).  Recently,  there has  been an attempt  to associate  the {\em
bottom}  magnetic field  of  the millisecond  pulsars  with the  total
amount  of mass  accreted \cite{zhan04}.   However, we  feel  that the
minimum field,  characteristic of  pulsars processed in  LMXBs, arises
due to an entirely different reason.  Because of the long evolutionary
time-scales the total amount of mass accreted by a neutron stars in an
LMXB  is quite large.   Addition of  this extra  mass may  trigger the
conversion of  a neutron  star into a  strange star  via deconfinement
\cite{olin87}.  Interestingly,  many of the  compact objects suspected
to be strange stars are residing in LMXBs \cite{li99,datt00}.  In view
of this,  we suggest that  it is the  intrinsic property of  a strange
star which gives rise to a lower-bound of the magnetic field.

{\em Strange Quark  Matter} (SQM), composed of u, d  and s quarks, may
probably be the ultimate ground state of matter \cite{witt84}.  It has
been  found that  the stable  range of  mass ($1\msun  -  2\msun$) for
strange stars is quite similar to that for neutron stars. Furthermore,
in this range  the radii of strange stars are  not very different from
those of the standard neutron stars \cite{haen86}.  Since the range of
stable rotation  periods sustainable by  these two types of  stars are
also similar,  it has always been  suspected that some  of the pulsars
could very  well be  strange stars \cite{alck86}.  In this  article we
take the  view that  if not all  pulsars then at least  the millisecond
pulsars could be strange stars.

\bef
\begin{minipage}{8.0cm}
\epsfig{file=fig01.ps,width=175pt,angle=-90}
\end{minipage}
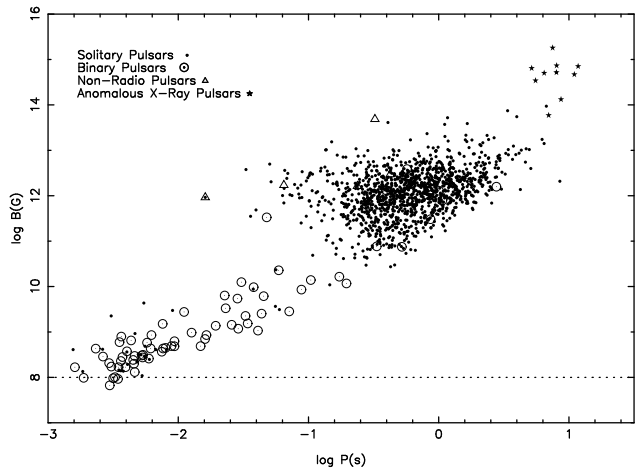
\caption[]{Magnetic field vs. spin-period of observed pulsars. The 
 data is obtained from the ATNF on-line pulsar catalog available at
 -- http://www.atnf.csiro.au/research/pulsar/psrcat/ and include
both galactic and extra-galactic pulsars.}
\label{fig01}
\eef

The aim  of the present  work is to  show that the physics  of strange
stars naturally gives rise to a  lower limit of the magnetic field. In
fact, a tuning  of the QCD parameters does produce  a value very close
to $10^8$~Gauss.  The paper is  organized as follows.  In section 2 we
discuss  our model  and describe  the calculations  in section  3. And
finally, we present our conclusions in section 4.

\section{di-quark formation}

To  explain the  observed  minimum  of the  pulsar  magnetic field  we
consider a rotating strange star model. The strange star is understood
to consist  of a $uds$ plasma  with a small admixture  of electrons to
maintain  overall charge  neutrality, where  each particle  species is
Fermi-degenerate~\cite{witt84,benv91a,benv91b}.   The  star  may  also
support   a   thin   hadronic   crust   of   mass   $M_{crust}   \lsim
10^{-5}$~\msun~\cite{alck86}. In  our discussion we  shall neglect the
effect of this crust.

In this work, we construct  a rotating strange star using the recently
proposed realistic quark  matter EOS in which --- (a)  a mean field is
derived  from a  two-body potential  incorporating  asymptotic freedom
with a deconfinement transition and (b) a density dependent ansatz for
the  quark mass  using  chiral symmetry  restoration  at high  density
\cite{deym98}.  This  EOS produces an  absolutely stable SQM  with the
same parameters for which $ud$  matter is unbound. The surface of such
a star is rather sharp,  since the deconfinement transition for strong
interaction sets in very quickly  at a critical density of about $\sim
~4.5 \rho_0$,  where $\rho_0  = 0.17 ~  /{\rm fm}^3$.  This  model has
been   successfully   used   to   describe  rotating   strange   stars
\cite{gond00,bomb00}.

Apart from  the mean field there  exists a spin-dependent  part of the
interaction  responsible,  for  example,  for the  $\pi-\rho$  or  the
$N-\Delta$ mass  splitting.  This interaction  potential, allowing the
quarks  to interact  with each  other  to form  di-quarks in  definite
color-spin channel, is given by~\cite{sinh02},
\beq 
V_{ij} = -  V_0(\lambda_i.\lambda_j) \, (S_i.S_j) \, e^{-\sigma^2
r_{ij}^2},
\label{eq_interaction}
\eeq
where  $V_0$  is  the  strength  and  $\sigma$ is  the  range  of  the
potential.  $\lambda$  and $S$  are the colour  and spin  matrices and
$r_{ij}$  is   the  distance  between   the  $i$-th  and   the  $j$-th
particle. $V_0$  and $\sigma$ are  adjusted such that  the predictions
for   $\pi-\rho$  or   $N-\Delta$  splitting   are  well   within  the
experimental  error  limits.   This   interaction  is  assumed  to  be
instanton induced with a  constant strength through the entire density
range.

There  can  be  two  types  of  di-quarks  ---  flavor  symmetric  and
anti-symmetric.  With  the negative sign  in Eq.(\ref{eq_interaction})
the  potential is  attractive in  two  combinations ---  (i) the  spin
singlet,  colour antisymmetric  state  ($\bar 3$)  and  (ii) the  spin
triplet,   colour   symmetric  state   (6).    With  such   spin-color
combinations, $ud$ pairs formed in the flavor antisymmetric state with
$L=0$  and $ss$  pairs formed  in the  $L=1$ state  will  decrease the
energy further \cite{sinh05}. Since the pairs are not physically bound
they would  experience stronger attraction  when they come  closer but
their kinetic  energy may overcome  the potential and take  them apart
again.  However,  on the  average there would be  a fixed
number  of  charged  di-quarks   at  any  instant.   Bailin  \&  Love
\citeyear{bail84}   have   shown   that   this  can   give   rise   to
superconductivity within  the star even when the  momentum transfer is
large.

In  a  rotating  star  with  one superconducting  species,  a  uniform
magnetic field (London field) is set up such that,
\beq 
B = \frac{2 m^* c}{q^*} {\Omega} \label{eq_london} 
\eeq
where $m^*$  and $q^*$ are the  effective mass and the charge  of the pair.
$\Omega$ is the angular velocity of the star \cite{baym88}.  Typically
the  strength  of  the London  field  is  rather  small and  is  $\sim
10^{-1}$~G for the $ud$ pair.

However, the  situation is much  more complicated with two  species of
di-quarks.   In a  rotating strange  star both  the $ud$  and  the $ss$
superconducting  pairs  are  present.    The  London  field  would  be
different  for  pairs  with  different  mass  and  charge.   From  the
minimization of the Ginzburg-Landau free energy Chau \citeyear{chau97}
concluded  that at  least  one species  will  rotate without  creating
vortices, when  the species  are non-interacting.  The  other species,
having a different London field,  would create vortices to rotate with
the same  angular velocity.  But  if the species are  interacting then
they would form vortex bundles surrounding a common normal core with a
magnetic  flux  trapped  within.   We  look  at  the  minimization  of
Ginzburg-Landau free energy  density of such a system  to estimate the
magnetic field of such a star.

\section{minimum energy state}

To  calculate the Ginzburg-Landau  free energy  of a  rotating strange
star, consisting of $ud$ and  $ss$ superconducting pairs, we assume the
pair momenta are small and the use of non-relativistic Ginzburg-Landau
theory is appropriate. Moreover, as mentioned above, we expect them to
exhibit  superfluidity. A  superfluid supports  rotation by  forming a
number of  Onsager-Feynman vortices each  carrying a fixed  quantum of
circulation inversely proportional to the effective mass.

For  a star  rotating with  an  angular velocity  ${\bf \Omega}$,  the
Ginzburg-Landau  free energy  functional  of a  di-quark assembly,  is
given by~\cite{saul89,chau97},  $F =  \int f \  dV$, where  the energy
density  $f$  consists   of  ---  (a)  the  magnetic   energy  of  the
superconductor,  (b) the  kinetic  and the  nucleation  energy of  the
vortices  and  (c)  the  interaction energy  between  different  quark
species if appropriate.  The order parameter for the $j$-th species is
$\Psi_j(\rb)$  ($=|\Psi_j| e^{i\phi_j(\rb)}$)  where  $\phi_j$ is  the
phase  and $r$  is the  distance variable.   Then, in  the co-rotating
frame, the Ginzburg-Landau free  energy density for the $j$-th species
can be written as,
\beq
f_j = -a_j{|\Psi_j|}^2
           + \frac{b_j}{2}{|\Psi_j|}^4
           + \frac{1}{2m_j^*} {|{\bf {\cal P}}_j \Psi_j|}^2
           + \frac{{|\nb \times \ab|}^2}{8\pi}
\label{eq_fed}
\eeq
where $a$ and $b$ are positive constants such that the superconducting
state is preferred over the  normal state and $m_j^*$ is the effective
mass of the  $j$-th pair. $\ab$ is the  corresponding vector potential
satisfying $\bb =  \nb \times \ab$. Because of  the rotation a uniform
magnetic field is set up such  that $\ab= (\bb \times \rb)/2$ along an
array of  vortices in the  superconducting interior of the  star.  The
kinetic  energy of  the  $j$-th species is determined  by its  momentum
operator ${\cal P}_j$, given by,
\beq 
{\bf {\cal P}}_j 
= - i \hbar \nb_j 
  + \frac{q^*_j}{c} \ab - m_j^*({\bf \Omega} \times \rb_j),
\label{eq_momentum}
\eeq
where  $q_j^*$ is  the effective  charge of  the pair.   If  the quark
species are non-interacting then the above-mentioned free energy takes
the following form~\cite{chau97}:
\ber
f 
&=& \frac{{\cal B}^2}{8 \pi} \nonumber \\
&& + \sum_j \left\{ \frac{\hbar \rho_{sj}}{m^*_j}
   \left[ \ln \left(\frac{R_{cj}}{\xi_j}\right) - \frac{3}{4} \right]
   + m^*_j \frac{\xi_j^2}{2\hbar} N_j(0) \Delta_j^2 \right\} \nonumber \\
&& \times \left| \Omega - \frac{q^*}{2 m^*_j c} {\cal B}\right|,
\eer
where,  $\rho_{sj}$ is  the  density of  the superconducting  species,
$R_{cj}$ is the inter-vortex  spacing, $\xi_j$ is the coherence length
or the radius of the core of the vortex and $\frac{1}{2}N(0) \Delta ^2
$ is the  difference in the energy density between  the normal and the
super-conducting phase.   The vortices of each of  the non-interacting
di-quark  species can  form its  own Abrikosov  lattice.   However, the
structure of $f$ is such that  there exists a situation where only one
species of  superconducting di-quarks  form vortices whereas  the other
species rotate without creating vortices.

However,  we  consider  the  more  realistic situation  in  which  the
superconducting  di-quark species  are  interacting.  The  interaction
induces  a drag  energy.  To  minimize this  drag a  vortex  bundle is
created  where all  the species  share  common normal  core of  radius
$\xi$.  Assuming the velocities of all the species to be the same, the
vortex bundle density $\cal D$ is given by :
\beq 
{\cal D} = 2 \Omega /K 
\label{vortex density} 
\eeq
where $K$ is the circulation constant defined as,
\beq 
K 
= \frac{h}{m^*_s q^*_{ud} - m^*_{ud} q^*_s} (q^*_{ud} N_s - q^*_s N_{ud}) \\
\label{K} 
\eeq
with  $N_j$ being  the number  of  vortex quantum  per bundle.   Then,
inclusion  of interaction changes  the free  energy functional  to the
following,
\ber
f 
&=& \frac{{\cal B}^2}{8 \pi} \nonumber \\
&+& \Big[\sum_j \left\{ \frac{h^2 \rho_{sj}N_j^2}{4 \pi m^{*2}_j}
   \left[ \ln \left(\frac{R_{cj}}{\xi_j}\right) - \frac{3}{4} \right]
   + \frac{\pi \xi_j^2}{2} N_j(0) \Delta_j^2 \right\} \nonumber \\
&& + \frac{\Phi_v^2}{8 \pi^2 \lambda^2} \Big] \times | {\cal D}|,
\eer
where $\Phi_v$  is the magnetic  flux in the  core of a  vortex bundle
given by,
\beq
\Phi_v 
= \frac{hc}{m^*_s q^*_{ud} - m^*_{ud} q^*_s} (m^*_{ud} N_s - m^*_s N_{ud}).
\label{Phi}
\eeq  

\section{minimum magnetic field}

Typically,  we  expect  the   system  to  favour  the  minimum  energy
configuration.   As can  be  seen from  the  form of  the free  energy
functional,  its minimization essentially  puts a  lower bound  on the
magnetic  field. Therefore,  once  the system  gets  locked into  this
minimum  energy state,  there  is  no natural  way  of decreasing  the
magnetic field further.

For the ground state configuration $N_{ud}=0$ and $N_s=1$, giving 
\beq 
K = \frac{h q^*{ud}}{m^*_s q^*_{ud} - m^*_{ud} q^*_s}, \; \;
\Phi_v = \frac{hc m^*_{ud}}{m^*_s q^*_{ud} - m^*_{ud} q^*_s}.
\eeq
Although there  are no vortex quanta  for the $ud$-pair  both the mass
and the charge of this pair enter  in $K$ as well as in $\Phi_v$. This
happens because of the strongly interacting nature of the system.

To  estimate  the  minimum   field  we  use  the  following  parameter
values. The  masses of  the di-quark pairs  are taken to  be $m^*_{ud}
\sim 270$~MeV  and $m^*_s \sim  560$~MeV \cite{sinh02}. With  these we
have, ${\cal  D } = 3.7  \times 10^3~\nu$~cm$^{-2}$ and  $\Phi_v = 3.4
\times 10^{-8}$~G~cm$^2$, where $\nu$ is the rotation frequency of the
star.  The  radial dimension of  the vortices, approximately  equal to
the inter-vortex distance, is obtained as follows,
\ber
R_c ~=~\frac{1}{\sqrt{\cal D }}~=~ \frac{0.016}{\sqrt{\nu}}~\mbox{cm}.\nonumber
\eer
For  example,  for  a  pulsar  with $P  =  1$~ms,  $R_c~=~2.02  \times
10^{-4}$~cm.  
The  penetration
depth $\lambda$ is given by,
\beq 
\lambda~=~\left(\frac{m^*_s c^2}{4 \pi~ n^*_s q^{*2}_s}\right)^{1/2}, 
\eeq
where  $n^{*}_s$  is  the  number  density  of  the  $ss$-pairs.   For
$ss$-pairs the  potential is attractive in the  $L=1$ channel implying
that the  centrifugal barrier in  the $p$-state reduces  the potential
effectively.  A simple  minded  calculation shows  that the  $ss$-pair
density is  $\sim 10^{-8}$~fm$^{-3}$ giving  rise to a large  value of
$\lambda$.  Using all these we have,
\ber
B = \frac{\Phi_v}{\pi \lambda^2} \sim  10^{8} \mbox{G}.
\eer
Evidently,  the small value  of $B$  is a  manifestation of  the large
$\lambda$.  

\section{conclusion}

In this  work, we have estimated  the minimum magnetic  field at which
the Ginzburg-Landau free energy of the SQM (with two types of di-quark
condensate) is minimized. For  suitable QCD parameters this appears to
be $\sim 10^8$~G.  It is mostly due to a large coherence length of the
$ss$-pair  ($\sim 10^5$~fm)  since they  are in  the $L=1$  state. The
centrifugal  barrier of  higher $L$-state  would reduce  the $ss$-pair
density. This and the large  strange quark mass increase the coherence
length,   $\xi$.  It  is   still  much   smaller  than   the  electron
superconductor  coherence length  but  it is  large  compared to  the
relevant QCD scale of 1~fm.

Although the  magnetic field of  $\sim 10^8$~G is  astronomically very
appealing,  given the uncertainties  in our  parameters and  the rough
nature of  the estimate the actual  value could differ by  an order of
magnitude or  so.  The  important conclusion of  our study  however is
obtaining a  lower bound  for the magnetic  field. Once this  value is
reached  the magnetic field  would not  spontaneously decay  (by ohmic
dissipation etc..) to  a lower field strength if  there is no external
perturbation present.  Therefore, if the neutron stars in the low mass
binary systems do indeed undergo a deconfinement transition to convert
into strange stars  we should not expect to see  any radio pulsar with
field values significantly smaller than $\sim 10^8$~Gauss.

Whether or  not some (or all) of  the pulsars are strange  stars is an
old debate. One  of the arguments against the  strange stars have been
the fact that  they would not be able to  sustain glitches because the
matter content of  the hadronic crust of such  stars are tiny compared
to the  magnitude of  the glitches observed  (see \cite{mads04}  for a
recent  update  on this  controversy).  The  recent  observation of  a
micro-glitch ($\delta P/P \sim 10^{-11}$) in the millisecond pulsar PSR
B1821-24 by Cognard \& Backer \citeyear{cogn04} lends some credence to
our hypothesis that the millisecond pulsars are strange stars. Because
a small hadronic  crust of a strange star would be  just right to give
rise to such a micro-glitch.

\section{acknowledgment}

We  would  like  to  thank  R.  Dey,  R.   Gopakumar,  B.~P.   Mandal,
T.~V. Ramakrishnan, S.   Ray and D. Shah for  helpful discussions. MD,
JD, MB acknowledge the hospitality of HRI, Allahabad and SK thanks the
astrophysics group of SISSA/ISAS, Trieste for the same.  This work has
been   supported  by   a  DST   (Government  of   India)   grant  (No.
SP/S2/K-03/01).

\bibliography{mnrasmnemonic,references}

\bibliographystyle{mnras}

\bsp

\label{lastpage}
\end{document}